\documentclass[a4paper,10pt,reqno,superscriptaddress]{revtex4}

\usepackage[centertags]{amsmath}
\usepackage{amsfonts}
\usepackage{amssymb}
\usepackage{newlfont}
\usepackage{stmaryrd}
\usepackage{mathrsfs}
\usepackage{euscript}

\usepackage{color}

\let\al=\alpha \let\be=\beta \let\de=\delta \let\ep=\epsilon
  \let\ga=\gamma 
  \let\om=\omega 
\let\si=\sigma

\let\De=\Delta

\newcommand{\caD}{{\mathcal D}}

\newcommand{\caL}{{\mathcal L}}

\newcommand{\caQ}{{\mathcal Q}}

\newcommand{\caS}{{\mathcal S}}
\newcommand{\caT}{{\mathcal T}}

\newcommand{\opunit}{\text{1}\kern-0.22em\text{l}}

\newcommand{\bsP}{{\boldsymbol P}}

\newcommand{\scD}{{\mathscr D}}

\DeclareMathAlphabet{\mathpzc}{OT1}{pzc}{m}{it}

\newcommand{\id}{\textrm{d}}

\begin{document}


\title{Steady state statistics of \\driven diffusions}

\author{Christian Maes}
\affiliation{Instituut voor Theoretische Fysica, K.U.Leuven, Belgium}
\email{christian.maes@fys.kuleuven.be}
\author{Karel Neto\v{c}n\'{y}}
\affiliation{Institute of Physics AS CR, Prague, Czech Republic}
\email{netocny@fzu.cz}
\author{Bram Wynants}
\affiliation{Instituut voor Theoretische Fysica, K.U.Leuven, Belgium}
\email{bram.wynants@fys.kuleuven.be}

\keywords{diffusion, nonequilibrium steady state, dynamical fluctuations}

\begin{abstract}
We consider overdamped diffusion processes driven out of thermal
equilibrium and we analyze their dynamical steady fluctuations. We
discuss the thermodynamic interpretation of the joint fluctuations
of occupation times and currents; they incorporate respectively
the time-symmetric and the time-antisymmetric sector of the
fluctuations. We highlight the canonical structure of the joint
fluctuations. The novel concept of traffic complements the entropy
production for the study of the occupation statistics. We explain
how the occupation and current fluctuations get mutually coupled
out of equilibrium. Their decoupling close-to-equilibrium explains
the validity of entropy production principles.
\end{abstract}

\maketitle

\section{Introduction}

Thermodynamics already ceases to be exact under the microscope.
That line of thought appeared a century ago, explicitly so in the
introduction to Einstein's paper on the atomic hypothesis,
\cite{Einstein1905}. The ensuing theory of Brownian motion started
a dynamical fluctuation theory, and essentially the same
mathematical model was further used by Onsager and Machlup,
\cite{ons}, to describe small macroscopic fluctuations around the
relaxation to equilibrium. It has developed into the (sometimes
called Lagrangian) approach to nonequilibrium statistical
mechanics which is based on giving a weight to available
trajectories in terms of an action that integrates a space-time
local Lagrangian. Since then various methods have been developed
to extract from that action functional physically relevant
information about the steady state statistics.

In recent years there has been a revival of nonequilibrium
fluctuation theory, with many contributions on various aspects of
the problem. One class are static fluctuations referring to the
statistics of macroscopic fluctuations upon a nonequilibrium
stationary distribution; they have been mostly studied for driven
lattice gases in the hydrodynamic regime, \cite{jona1}, by
generalizing the Hamilton-Jacobi method originally proposed
in~\cite{kubo}. Related yet fundamentally different are the
dynamical fluctuations concerning time-integrated observables.
That is what this paper is about.  We will introduce and also
compare our present work with other recent approaches in Section
\ref{compa}.  There has been a particular interest in the
(integrated) current and entropy production fluctuations, also in
the context of the celebrated fluctuation symmetry. It is nearly
impossible to mention all the relevant references; as an example
we include~\cite{ecv,GC,jona,bd,der,geneve,FNJ}.

An extension of the dynamical approach to the time-symmetric
domain was initiated in~\cite{maar,woj,dv1}. In particular, it has
been shown how the nonequilibrium fluctuations of the steady state
occupation times are related to the mean entropy production,
\cite{dv1}. That relation proves useful for understanding the
status and the limitations of entropy production principles in
characterizing the steady state, \cite{dv1,dv2}. As further
unfolded in~\cite{mn3} within the framework of Markov jump
processes, a particularly simple and generic structure of
dynamical fluctuations emerges when both the time-symmetric
(occupation times) and the time-antisymmetric (current)
fluctuations are observed jointly. Their correlation can be
neglected in a close to equilibrium regime, where the familiar
linear irreversible scheme based on the entropy production alone
exists. However, the coupling between the time-symmetric and
time-antisymmetric sectors becomes essential far from equilibrium.
That can be seen as a fundamental reason for highly complex
patterns of e.g.\ current fluctuations which arise when observing
their marginal distribution alone. The main question remains to
see some systematics in the fluctuation formul\ae\ and to identify
terms in the rate functions that have a general thermodynamic
meaning. In order to complete that program the following examples
will help as orientation.

The plan of the paper is as follows. In Section~\ref{sec: model}
we introduce a class of driven diffusion models on which we
formulate our main questions and answers. In Section~\ref{sec:
thermodynamic} we define the dynamical entropy, composed of a
time-antisymmetric part (entropy flux) and a time-symmetric part
(dynamical activity or traffic). Our main results come in
Section~\ref{sec: fluctuations} where we show how entropy
production functionals govern the dynamical fluctuations of
diffusion processes. We see then how the time-symmetric and
the time-antisymmetric fluctuations get coupled, and we discuss
the validity of some entropy production principles. The
mathematics involved is the standard It\^o/Stratonovich stochastic
calculus complemented with the path-integral formalism; for convenience we
give a brief review in the appendix together with some additional
references.

\section{Model and main results}\label{sec: model}

We start with an example of a one-dimensional overdamped Langevin dynamics driven from equilibrium by a nongradient force. This is further generalized
to a larger class of diffusions, and in arbitrary dimension.

\subsection{Basic example: diffusion on a circle}\label{sec: circle}

We consider a particle undergoing an overdamped motion on the
 circle with unit length. This means that we ignore inertial effects, so that forces will be proportional to velocities rather than to acceleration. The particle moves under the influence of a stochastic force (noise), because we imagine the system to be connected to an environment at inverse temperature $\beta$. Added to that, there are also deterministic forces: a periodic potential
landscape $U(x)$, and a periodic (nongradient) force $F(x)$, which
will drive the system out of equilibrium. Note that on the circle
every function $F$ is periodic, but is not in general a derivative
of a (periodic) function (for example if $F > 0$ everywhere). With
some (possibly inhomogeneous) mobility $\chi(x) > 0$, this is
modeled by the It\^o-stochastic dynamics (see Appendix A):
\begin{equation}\label{gsd}
  \id x_t = \chi(x_t)\big[ F(x_t) - U'(x_t)\big] \id t + D'(x_t) \id t
  + \sqrt{2D(x_t)}\, \id B_t
\end{equation}
where $\id B_t$ is standard Gaussian white noise, so that the process is Markov.
 The prime as
 superscript is a shorthand for the spatial derivative. The
 diffusion coefficient is $D = \chi / \be$, in agreement with the condition of local
detailed balance. The drift counter-term proportional to $D'$ then
ensures that the case $F = 0$ is an equilibrium dynamics, see
below. For the following we assume that this stochastic process
\eqref{gsd} always relaxes to a unique stationary distribution
which mathematically amounts here to smoothness conditions on
$F,U$ and $\chi$, see e.g. in \cite{dif}.

The corresponding Fokker-Planck equation for the time-dependent probability density
$\mu_t$ is
\begin{equation}\label{fk}
 \frac{\partial\mu_t(x)}{\partial t} + j_{\mu_t}'(x) = 0,\qquad
 j_\mu = \chi \mu (F-U') -  D\mu'
\end{equation}
where $j_\mu$ is the probability current as the sum of a drift and a diffusion component.

The stationary density, called $\rho$ in the sequel, solves the
stationary equation $j'_\rho = 0$,
i.e.,
\begin{equation}\label{circle-stat}
  \chi \rho (F - U') - D \rho' = j_\rho
\end{equation}
is a constant. For $F = 0$, equation~\eqref{circle-stat} has the solution
\begin{equation}\label{bg}
  \rho(x) = \frac 1{Z}\,e^{-\be U(x)},\qquad
  Z = \int_0^1 e^{ -\be U} \id x
\end{equation}
and the corresponding stationary current is $j_\rho = 0$; this is
a detailed balanced dynamics
with $\rho$ the equilibrium density.\\
When adding a nongradient driving force,
$\int_0^1 F \id x \neq 0$, the stationary density obtains the form
\begin{equation}
  \rho(x) = \frac{1}{\cal Z}
  \int_0^1 \frac{e^{\be W(y,x)}}{D(y)}\,
   \id y,\qquad {\cal Z} =
    \int_0^1\int_0^1 \frac{e^{\be W(y,x)}}{D(y)}\, \id y\id x
\end{equation}
where
\begin{equation}
  W(y,x) = U(y) - U(x) +
  \begin{cases}
    \int_y^x F\, \id z  &  \text{for } y \leq x
  \\
    \int_y^1 F\, \id z + \int_0^x F \id z  &  \text{for } y > x
  \end{cases}
\end{equation}
is the work performed by the applied forces along the positively
oriented path $y \rightarrow x$. In this model the stationary
current can be computed by dividing the stationary
equation~\eqref{circle-stat} by $\rho\chi$ and by integration over
the circle:
\begin{equation}\label{circle-current}
  j_\rho = \frac{\overline{W}}{\int_0^1 (\rho\chi)^{-1}\id x}
\end{equation}
where $\overline{W} = \int_0^1 F \id x$ is the work carried over a completed cycle.
The non-zero value of this stationary current indicates that time-reversibility is broken.
In the simplest nonequilibrium setting when $U = 0$ and $F, \chi > 0$ are some constants, the
steady state has the uniform density $\rho(x) = 1$ and the current is $j_\rho = \chi F$.

\subsection{General model}\label{sec: general model}

As a generalization of the driven diffusion on a circle, we consider a class of
$d-$dimensional inhomogeneous diffusions introduced by the following equation,
which has to be interpreted in the It\^o way (see Appendix A):
\begin{equation}\label{gsd1}
  \id x_t = \bigl\{\chi(x_t) \bigl[ F(x_t) -
   \nabla U(x_t) \bigr] + \nabla\cdot D(x_t)\bigr\}\,\id t
  + \sqrt{2 D(x_t)}\, \id B_t
\end{equation}
The mobility $\chi(x)$ is now a strictly positive (symmetric) $d
\times d-$matrix which is related to the diffusion matrix $D(x)$
as $\chi(x) = \be D(x)$ with some fixed homogeneous inverse
temperature $\be > 0$, and the $d-$dimensional vector $\id B_t$
has independent standard Gaussian white noise components. The
notation $\nabla \cdot D$ stands for the vector with components
$\sum_\ga \partial_\ga D_{\ga \al}$, i.e., with the derivative
acting on the left indices of a matrix. From now on we also use
the dot to denote scalar product, e.g., $F \cdot G = \sum_\ga
F_\ga G_\ga$, whereas no dot is used for a matrix acting on a
vector, e.g., $(\chi\, F)_\al = \sum_\ga \chi_{\al\ga} F_\ga$.

We restrict ourselves to two types of boundary conditions:\\
(1) \emph{periodic}---the particle moves on the unit torus
$[0,1)^d$ and the fields $U$, $F$,
and $\chi$ are smooth functions on the torus;\\
(2) \emph{decay at infinity}---the potential $U$ grows fast enough at infinity so that the
particle is essentially confined to a bounded
 region, i.e., the density and its derivative vanish at infinity.

Under either of the above boundary conditions we can simply ignore
boundary terms when performing integrations by parts.  The particles are essentially
 confined in their configuration space.\\

The probability density $\mu_t$ evolves according to the
Fokker-Planck equation associated with~\eqref{gsd1}:
\begin{equation}\label{fk1}
  \frac{\partial\mu_t}{\partial t} + \nabla\cdot j_{\mu_t} = 0,\qquad
  j_\mu = \chi \mu \,(F - \nabla U) - D\nabla\mu
\end{equation}
where again $\chi = \beta D$. The stationary condition reads
$\nabla \cdot j_\rho = 0$; in contrast to the previous
one-dimensional example the stationary current in general becomes
inhomogeneous here. Equation \eqref{fk1} can be interpreted as
giving the evolution of the density profile $\mu_t$ of a
macroscopic amount of independent particles each moving according
to~\eqref{gsd1}: at times $t\geq 0$ and for all observables $f$,
\begin{equation}\label{str-identity0}
  \Bigl\langle  f(x_t) \Bigr\rangle_{\mu_0}
  = \int f(x)\,\mu_t(x)\,\id x
\end{equation}
where $\langle\cdot\rangle_{\mu_0}$ is the average over the
path-space distribution started at $\mu_0$.\\
Moreover, the $j_\mu$ then also gives the expected profile of the
`real' particle current at given density $\mu$: again under the
diffusion process started from density~$\mu_0$,
\begin{equation} \label{str-identity}
  \Bigl\langle \int_0^T f(x_t) \circ \id x_t \Bigr\rangle_{\mu_0}
  = \int_0^T \id t \int f(x)\, j_{\mu_t}(x)\,\id x
\end{equation}
which involves the Stratonovich-stochastic integral; see
\eqref{a8} in Appendix \ref{stin} for its derivation. Taking
formally $f(\,\cdot\,) = \de(\,\cdot\,-x)$, we indeed recognize
$\mu_t$ and $j_{\mu_t}$ as the time-dependent (or transient) local
density, respectively the local current for that density.

\subsection{Questions and first answers}\label{compa}

Since the Langevin dynamics~\eqref{gsd} or \eqref{gsd1} is assumed
to have a unique stationary distribution, the stationary density
$\rho$ can be measured as the most typical time-average of the
occupation over a large time interval. Any different measurement
outcome, specified by some density $\mu$, is well possible but it
is a rare event with an increasingly small probability to be
observed as the duration $T\uparrow +\infty$. The large deviation
theory, \cite{DV,DZ}, tells us that the generic asymptotic law for
such fluctuations is an exponential decay, $\bsP_T(\mu) \sim e^{-T
I(\mu)}$ with some rate function $I(\mu)$. Here comes a first
question:

\textbf{QA}: How can we find the rate function $I(\mu)$ and what
is its thermodynamic meaning?

It turns out that $I(\mu)$ can be expressed in terms of the
\emph{traffic} measuring the dynamical activity in the system, and
being introduced as a time-symmetric counterpart to the entropy
production. We also give a relation between the traffic and the
entropy production which appears to be general for diffusion
processes.\\
A very similar question can also be asked about the current
statistics, with rate function $I(j)$, but the study of
$I(\mu)$ is quite new.\\

\textbf{QB}: What are the fluctuations of the time-averaged
current around its most probable value $j_\rho$, and how are they
correlated with the fluctuations of the occupation times?  In
particular, what are the joint fluctuations for both
time-symmetric and time-antisymmetric observables?

To answer these questions, we give an explicit form of the
dynamical fluctuation functional $I(\mu,j)$ such that $\bsP(\mu,j)
\sim e^{-T I(\mu,j)}$ is the joint probability to see both $\mu$
as the statistics of occupations times and $j$ as the
time-averaged current. The functional $I(\mu,j)$ has a general
canonical structure and it is given in terms of the entropy
production and the traffic. Starting from this basic functional
other fluctuation formulas can be obtained by the contraction
principle familiar from large deviation theory. For example, one
recovers $I(\mu)$ by minimizing $I(\mu,j)$ over all admissible
currents $j$.\\

The above questions and the methods that we take below are not
entirely original.  They have appeared in the mathematical
literature in a systematic way since the theory of large
deviations was introduced in the framework of Markov diffusion
processes, see \cite{DV,DZ,var}. The relevance to physics and to
statistical mechanics in particular is obvious, but the
thermodynamic interpretation of the resulting dynamical
fluctuation functionals has not been systematically investigated.
A first study can be found in \cite{mn3}. Our paper will add to
that, starting from the next section. There have of course been
many other studies of dynamical fluctuation theory in the
literature. We mention in particular the works of Derrida and
Bodineau, see \cite{bd,der} and of Bertini {\it et al}, e.g. in
\cite{jona1,jona}. As we are dealing here with diffusion
processes, our approach is especially similar to what one is doing
for a macrostatistical theory, where the hydrodynamic fluctuations
can be viewed as solution of some infinite-dimensional diffusion
process.  Possible differences with the existing work are first of
all that the problems related to the diffusion-approximation or to
a hydrodynamic
 rescaling do not enter in our work.  We just start from a
  finite-dimensional diffusion process
 as such, in the same way  as we
 could start from a Markov jump process, boundary or bulk driven, and without extra
 rescaling.  In other words, we prefer to split the problem of
 hydrodynamical scaling with possible diffusion approximation from
 the problem of studying the dynamical fluctuations.  Secondly, we
 are concerned here directly with the stationary fluctuations,
 without passing via the transient regime.
 Thirdly, as soon will become clear, we like to emphasize a (thermodynamic)
 canonical structure in the joint fluctuations of density and
current.  In order to achieve that,
 it is natural to introduce the novel concept of traffic (as in the next section)
 whose thermodynamic
 interpretation remains to be fully elucidated.  Our three main
 results are in Sections \ref{res1}, \ref{res21} and \ref{occ}.

\section{Thermodynamic considerations}\label{sec: thermodynamic}

In this section we introduce the two main players in the presented
fluctuation theory: the entropy production and the traffic. While
the former notion is rather standard, the latter appears to be
new.

\subsection{Entropy production}

The entropy production is the change of entropy in the world
between two moments in a process. The very notion of entropy is
however controversial when trying to extend it to nonequilibrium
situations.  The problem is mainly that we do not have a good
understanding of the relevant quantities or of the macroscopic
variables that characterize the nonequilibrium state and its
evolution. Nevertheless we can try to work with the notion of
entropy production as it has come to us from considerations of
heterogeneous or of local equilibrium.  There are two basic
contributions to that entropy production.

The entropy of a closed macroscopic system is defined via counting
microstates, which at the same time gives both the static
fluctuation theory with the entropy as a fluctuation functional
and a second law inequality, \cite{Hthm}. For a mesoscopic system
this cannot be applied as such. However, one can think of an
ensemble of $N \to \infty$ independent copies. A density $\mu$
then becomes a macro-observable telling us the relative occupation
of the space, and the associated counting entropy equals the
relative entropy (with respect to the flat distribution):
\begin{equation}\label{static-ent}
  s(\mu) = -\int \mu(x) \log \mu(x)\,\id x
\end{equation}
By construction it is the (static) fluctuation functional in the
probability law for observing the empirical density $\mu$ when
sampling the particles from the flat distribution. For $F = \nabla
U = 0$ (i.e.\ for a thermodynamically closed system) the entropy
rate satisfies $\id s / \id t \geq 0$ along the solution of the
Fokker-Planck equation.

An open system dissipates heat that results in an extra entropy
production in the environment. The rate $\caQ(\mu)$ of mean heat
dissipation comes from the work performed by the nongradient force
and from the change of the energy of the system:
\begin{equation}
\begin{split}
  \caQ(\mu) &= \int F \cdot j_\mu\,\id x
  - \frac{\id}{\id t} \int U\,\mu\, \id x
\\
  &= \int (F - \nabla U) \cdot j_\mu\,\id x + \int \nabla \cdot (U j_\mu)
  \,\id x
\end{split}
\end{equation}
where we have used the Fokker-Planck equation~\eqref{fk1}. Under
either of the two boundary conditions considered in this paper
(i.e.\ periodic or decaying at infinity, see Section~\ref{sec:
general model}), the last integral equals zero. That argument will
often come back in what follows.

The environment is a heat reservoir at inverse temperature $\be$
that remains itself at equilibrium during the whole process (the
weak coupling assumption). Hence, the mean entropy flux equals
$\be \caQ(\mu)$ and the total entropy production rate reads
\begin{equation}\label{ent-prod}
\begin{split}
  \si(\mu) &= \be \caQ(\mu) + \frac{\id s}{\id t}(\mu)
\\
  &= \int (\be F - \be\nabla U - \nabla \log \mu) \cdot j_\mu\,\id x
\\
  &= \int j_\mu \cdot (\mu D)^{-1} j_\mu\, \id x
\end{split}
\end{equation}
where we have used again the Fokker-Planck equation~\eqref{fk1}
and the boundary conditions. In the language of irreversible
thermodynamics, $\be F - \nabla \log (\mu\, e^{\be U})$ is a
generalized thermodynamic force conjugated to the current $j_\mu$.
Clearly, $\si(\mu) \geq 0$ in agreement with the second law,
proving the thermodynamic consistency of our diffusion model.

\subsection{Dynamical entropy and traffic}

The diffusion as defined by the stochastic equation~\eqref{gsd1}
is a Markov process. Its randomness can be characterized by the
dynamical entropy which is a dynamical variant of the static
entropy~\eqref{static-ent} obtained by replacing the density $\mu$
with the path-space distribution of the process. Such a
construction is often useful in the theory of dynamical and
stochastic systems; here we review how it is linked to the
thermodynamic entropy production defined in the previous section
and what other information it provides.

We start by writing the density of the path-distribution
$\bsP_{\mu_0}$ of the process with respect to a suitable reference
process $\bsP^0$, for which we take the one corresponding to $F =
\nabla U = 0$ (see Appendix B):
\begin{equation}\label{path-distribution}
  \id\bsP_{\mu_0}(\om) = \mu_0(x_0)\,e^{-\int_0^T \caL^+(x_t)\,\id t
  - \int_0^T \caL^-(x_t,\id x_t)}\id\bsP^0(\om)
\end{equation}
where $\om= (x_t)_{t=0}^T$ is a trajectory and $\mu_0$ is an
initial density. The weight in the exponential is split in a
time-symmetric part
\begin{align}\label{lagr-plus}
  \caL^+(x) &= \frac{\be^2}{4} [ (F - \nabla U) \cdot D(F - \nabla U) ]
  + \frac{\be}{2} \nabla \cdot [ D(F - \nabla U) ]
\\\intertext{and a time-antisymmetric part}
\label{lagr-minus}
  \caL^-(x,\id x) &= -\frac{\be}{2} (F - \nabla U) \circ \id x
\end{align}

The dynamical entropy is the relative entropy of the
path-distribution $\bsP_\mu$ with respect to the reference
$\bsP^0$, over some fixed time interval $[0,T]$:
\begin{equation}\label{dyn-ent}
\begin{split}
  \caD(\mu_0) &=  \Bigl\langle \log\frac{\id P_{\mu_0}}{\id P^0} \Bigr\rangle_{\mu_0}
  = -s(\mu_0) - \Bigl\langle \int_0^T \caL^+(x_t)\,\id t \Bigr\rangle_{\mu_0}
  - \Bigl\langle \int_0^T \caL^-(x_t,\id x_t) \Bigr\rangle_{\mu_0}
\end{split}
\end{equation}
The time-antisymmetric contribution to the dynamical entropy equals,
applying the identity~\eqref{str-identity},
\begin{equation}
  -\Bigl\langle \int_0^T \caL^-(x_t,\id x_t) \Bigr\rangle_{\mu_0}
  = \frac{\be}{2} \int_0^T \id t\int (F - \nabla U) j_{\mu_t}\,\id x
  = \frac{\be}{2} \int_0^T \caQ(\mu_t)\,\id t
\end{equation}
Hence, it is given in terms of the entropy flux. This is an
instance of a fairly general observation that lies behind a
famous fluctuation symmetry of the entropy production, \cite{mn,poincare,gav}.\\
 The
time-symmetric contribution is computed analogously
via~\eqref{str-identity0} to obtain
\begin{equation}
  \Bigl\langle \int_0^T \caL^+(x_t)\,\id t \Bigr\rangle_{\mu_0}
  = \int_0^T \caT(\mu_t)\,\id t
\end{equation}
where we have introduced a new quantity $\caT(\mu)$, called \emph{traffic}, which equals
\begin{equation}\label{traffic-def}
  \caT(\mu) = \frac{\be^2}{4} \int \mu(F - \nabla U) \cdot D(F - \nabla U)\,\id x
  + \frac{\be}{2} \int \mu \nabla \cdot [ D(F - \nabla U)]\,\id x
\end{equation}
By construction, the traffic is that part of the dynamical entropy
which originates from the time-symmetric fluctuations. Finally, we arrive at
the next two equivalent expressions for the dynamical entropy
\eqref{dyn-ent}:
\begin{equation}
\begin{split}
  \caD(\mu_0) &= -s(\mu_0) - \int_0^T \caT(\mu_t)\,\id t
  + \frac{\be}{2} \int_0^T \caQ(\mu_t)\,\id t
\\
  &= -\frac{1}{2} \bigl[ s(\mu_0) + s(\mu_T) + 2\int_0^T \caT(\mu_t)\,\id t
  - \int_0^T \si(\mu_t)\,\id t \bigr]
\end{split}
\end{equation}
where we have inserted \eqref{ent-prod}.\\

\subsection{Traffic versus entropy production}
A remarkable and important feature of diffusion systems, not valid in general
beyond the diffusion approximation, \cite{mn3}, is that the
traffic~\eqref{traffic-def} and the entropy
production~\eqref{ent-prod} are not independent of each other.
Indeed, one checks the relation
\begin{equation}\label{ent-traffic}
  \caT(\mu) = \frac{\si(\mu)}{4}
  + \int \frac{\nabla\mu \cdot D \nabla\mu}{4\mu}\,\id x
\end{equation}
in which the last term only depends on the distribution $\mu$ and
not on the imposed potential $U$ nor on the driving $F$. As we
will see in the next section, differences in the traffic and in
the entropy production when varying $U$ and $F$ determine the
asymptotics of dynamical fluctuations. Hence, relation
\eqref{ent-traffic} brings about a simplification in the structure
of fluctuations that is characteristic and restricted to
diffusions.  We believe that this also indicates that a more
general nonequilibrium theory should also reach beyond the
Langevin or diffusion approximation.



\section{Joint occupation-current statistics}\label{sec: fluctuations}

According to Einstein's fluctuation theory, the entropy and
derived quantities govern the structure of static fluctuations. In
this section we explain how the time-symmetric and
time-antisymmetric components of the dynamical entropy play an
essential role in dynamical fluctuation theory.

\subsection{Definitions}
 A basic and time-symmetric dynamical
observable is the empirical density of the occupation times
defined as the fraction of time spent in the neighborhood of every
point $x$, over a fixed time interval $T$:
\begin{equation}
  \bar\mu_T(x) = \frac{1}{T} \int_0^T \de(x_t - x)\,\id t
\end{equation}
This is a path-dependent observable as it varies over the paths
$\omega = (x_t)_{t=0}^T$.  For the large time asymptotics $T \to
\infty$, we have $\bar\mu_T \to \rho$ almost surely and
independently of the initial condition. We will be concerned with
physically determining the fluctuation functional $I(\mu)$ that
enters the large deviation law
\begin{equation}\label{ocup-LD}
  \bsP(\bar\mu_T = \mu) \sim e^{-T I(\mu)}
\end{equation}
That has to be understood as an asymptotic formula that becomes an
equality after taking the logarithm and dividing by $T \to \infty$
in both left- and right-hand sides; see~\cite{rem,DZ} for a more
precise mathematical formulation.\\

The time-antisymmetric observable of special relevance is the
empirical current, formally defined as
\begin{equation}
  \bar j_T(x) = \frac {1}{T} \int_0^T \de(x_t - x) \circ \id x_t
\end{equation}
It depends again on the (random) path $\omega = (x_t)_{t=0}^T$ and
it measures the time-averaged current while in $x$, as in
\eqref{str-identity}.  Its steady average equals $j_\rho$;
moreover, $\bar j_T \to j_\rho$ for $T \to \infty$ almost surely.
The rate function $I(j)$ is defined analogously to
\eqref{ocup-LD}, but for the probability $\bsP(  \bar j_T = j)$.\\

Both $I(\mu)$ and $I(j)$ can be obtained in principle from the
joint fluctuations.
 The large time asymptotics
of the joint fluctuations of $\bar\mu_T$ and $\bar j_T$ is
described by the fluctuation functional $I(\mu,j)$ such that
\begin{equation}\label{imuj}
  \bsP(\bar \mu_T = \mu;\,\bar j_T = j) \sim e^{-T I(\mu,j)}
\end{equation}
always logarithmically when $T \to \infty$. A first observation is
that $I(\mu,j) = \infty = I(j)$ whenever $j$ is not stationary,
i.e.\ for $\nabla \cdot j \neq 0$. Indeed, for any smooth bounded
function $Y$ one has
\begin{equation}
\begin{split}
  \int Y\, \nabla \cdot \bar j_T\, \id x =
  -\frac{1}{T} \int_0^T \nabla Y(x_t) \circ \id x_t
  = -\frac{1}{T} [Y(x_T) - Y(x_0)] \to 0
\end{split}
\end{equation}
and hence, in a distributional sense, $\nabla \cdot \bar j_T \to
0$ for $T \to \infty$ along \emph{any} particle trajectory, which
proves the above statement. That is why from now on we always
assume that $\nabla \cdot j = 0$, unless otherwise specified.

\subsection{Result 1: traffic and entropy production
 determining the joint fluctuations}\label{res1}
 To compute $I(\mu,j)$ we use a standard large
deviation technique, sometimes referred to as Cram\'er tilting,
see e.g. \cite{var,DZ}.
 We modify
 the driving of the original dynamics changing $F$ into some $G$ and we take care that
 this new and modified dynamics is chosen  so that $\mu$ and $j$ become both
stationary (and hence are typical as large time averages).
  The potential $U$ remains
untouched. The more formal argument goes as follows.\\
Let $\bsP_F$  and $\bsP_G$ be the stationary path-space
distributions for driving $F$ respectively $G$, both processes
started at their respective stationary distribution. We refer to
Appendix \ref{discretization} for that notion of path-space
distribution.
 We must compute the probability $\bsP_F[A]$ of the event $A$ containing all the
 paths $\omega$
 over time $T$ such that $\bar \mu_T = \mu$ and $\bar j_T = j$, see
 \eqref{imuj}:
 \begin{equation}\label{cram}
\bsP_F[A] = \int \id \bsP_G(\omega)\,\frac{\id \bsP_F}{\id
\bsP_G}(\omega)\, \chi[\bar \mu_T = \mu;\,\bar j_T = j]
\end{equation}
The density of path-space distribution $\bsP_F$ with respect to
path-space distribution $\bsP_G$ can be written out explicitly
starting from \eqref{b3}:  we find
\[
\frac{\id \bsP_F}{\id \bsP_G}(\omega) =
\frac{\rho(x_0)}{\mu(x_0)}\exp\big[\int_0^T \id
t\,(\cal{L}^+_G-\cal{L}^+_F + \cal{L}^-_G - \cal{L}^-_F)\big]
\]
where we have made use of definitions \eqref{lagr-plus} and
\eqref{lagr-minus}.  The point is now that this can be
 fully
expressed as a function of the path-dependent occupation and
current fractions; in other words, when indeed $A$ occurs, that
is, when $\bar \mu_T = \mu$ and $\bar j_T = j$, then
\begin{align}\label{L+ mod1}
  \int_0^T \cal{L}^+_G(x_t)\,
  \id t &= T \int \mu(x) \cal{L}^+_G(x)\,\id x = T \caT_G(\mu)
\\
\label{L- mod1}
  \int_0^T \cal{L}^-_G(x_t,\id x_t)\, &=
   -\frac{\be}{2} \int \id x\, (G - \nabla U)(x) \cdot
  \int_0^T \de(x_t - x) \circ \id x_t
\nonumber
\\
  &= -\frac{\be T}{2} \int G \cdot j\,\id x
\end{align}
As a consequence \eqref{cram} simplifies: the density between the
path-space distributions comes out of the path-integral and since
by construction, for $T\rightarrow \infty$,  $\bsP_G[A]
\rightarrow 1$, we get \eqref{imuj} in the form
\begin{equation}\label{main0}
  I(\mu,j) = \caT_F(\mu) - \caT_G(\mu) +
  \frac{\be}{2} \int (G - F) \cdot j\,\,\id x
\end{equation}
That is to be read as the sum of an excess instantaneous traffic
given density $\mu$ alone, and an excess work (or, equivalently,
entropy flux) given stationary current $j$ alone. The excess is
being understood in the sense of the above modification, with $G$
producing the current $j$, i.e., $j = \chi\mu(G-\nabla U) -
D\nabla \mu$, see \eqref{fk1}, or
\begin{equation}\label{gee}
  \be G = (\mu D)^{-1} j + \nabla\log (\mu\, e^{\be U})
\end{equation}
Since, by \eqref{ent-traffic}, excess entropy production equals
excess traffic (for our diffusion processes), we can equivalently
write \eqref{main0} as
\begin{equation}\label{main1}
  I(\mu,j) = \frac{\si(\mu) - \si_G(\mu)}{4} + \frac{\be}{2} \int (G - F) \cdot j\,\,\id x
\end{equation}
Upon substituting \eqref{gee} into~\eqref{main1} and using
\eqref{ent-prod}, the fluctuation functional can also be written
in the following explicitly positive form:
\begin{equation}\label{main2}
\begin{split}
  I(\mu,j) &= \frac{\si(\mu) - \si_G(\mu)}{4}
  + \frac{1}{2} \int (\mu D)^{-1} (j - j_\mu) \cdot j\,\,\id x
\\
  &= \frac{1}{4} \int (j - j_\mu) \cdot (\mu D)^{-1} (j - j_\mu)\,\id x
\end{split}
\end{equation}
(On the assumption $\nabla \cdot j = 0$; remember that $I(\mu,j) =
\infty$ otherwise.) This last formula resembles the Gaussian-like
expressions for the current distribution, typical for hydrodynamic
fluctuations of the diffusion-type.  Such expressions are
omnipresent in the works of e.g. \cite{bd,jona}.  Although the
quadratic integrand in~\eqref{main2} resembles the (generalized
Onsager-Machlup) Lagrangian for macroscopic fluctuations in the
hydrodynamic limit, we have no spatial/temporal rescaling here. We
have started from a mesoscopic system as described by a diffusion
equation and the only large parameter is the time span $T$. What
we stress here with respect to other work on dynamical large
deviations, becomes visible from the formul\ae  \eqref{main0} and
\eqref{main1} and has as such nothing to do with hydrodynamic
rescaling or with macrostatistics.  It concerns the thermodynamic
interpretation of the fluctuation functional $I(\mu,j)$ for our
mesoscopic system: how it is shaped from quantities like traffic,
work and entropy production, and providing full account of the
steady dynamical fluctuations in both the time-symmetric and the
time-antisymmetric sectors.

\subsection{Steady state fluctuation symmetry}
As proof of internal consistency we check the fluctuation
symmetry, cf. \cite{ecv,GC,gav}.\\
 As a corollary
of~\eqref{main2}, one has
\begin{equation}\label{flu}
  I(\mu,-j) - I(\mu,j) = \int j \cdot (\mu D)^{-1} j_\mu \,\id x
  = \caS(j)
\end{equation}
where
\begin{align} \caS(j) =& \;\be \int F \cdot j\,\,\id
x\label{eflux}\\
=& \;\frac{\beta}{T}\int_0^T\, F(x_t) \circ \id x_t \quad
\text{(when $\bar j_T = j$)}\nonumber
\end{align}
 is $\beta$ times the time-averaged
power of the nongradient forces under the condition that $\bar j_T
= j$ is the time-averaged current. Due to the stationary condition
($\nabla \cdot j = 0$) it also coincides with the entropy flux.

\subsection{Result 2: canonical structure of the joint fluctuations}\label{res21}

 A useful feature of equilibrium
statistical thermodynamics is that the fluctuation functions can
in a simple and general way be given as differences of
thermodynamic potentials that generate the relevant order
parameters.
 As suggested already by formula~\eqref{main0}, a
similar structure also emerges for the dynamical fluctuations. In
order to make that manifest we should follow the exact dependence
on the driving $F$, and we therefore indicate that dependence here
explicitly.  In that way, we have an immediate rewriting
of~\eqref{main0}:
\begin{equation}\label{1rew}
 I_F(\mu,j) =  I_0(\mu,j)- \frac{\caS_F(j)}{2}+ \caT_F(\mu) - \caT_0(\mu)
 \end{equation}
where $I_0(\mu,j)$ is the fluctuation functional for the reference
equilibrium process with no driving ($F=0$), traffic was defined
in \eqref{traffic-def} and the entropy flux follows from
\eqref{eflux}. The canonical structure of that fluctuation
functional arises because the last difference, the excess traffic
\[ H(\mu,F) = 2\,[\caT_F(\mu) - \caT_0(\mu)]
 \]  is a potential
for the current in the sense that
\begin{equation}\label{currentpotential}
  \frac{\de H(\mu,F)}{\de F(x)}
   = 2\frac{\de \caT_F(\mu)}{\de F(x)} = \be \, j_\mu^F(x)
\end{equation}
Its Legendre transform is
\[
G(\mu,j) = \sup_F\, [\caS_F(j)  - H(\mu,F)]
\]
with associated
\begin{equation}\label{forcepotential}
  \frac{\de G}{\de j(x)}(\mu,j^F_\mu)
   = \beta\,F(x)
\end{equation}
so that force and current $(F,j)$ make a canonical pair. It can be
checked immediately that
\begin{equation}
 G(\mu,j) =
 \frac{1}{2} \int (j - j_\mu^0) \cdot (\mu D)^{-1} (j - j_\mu^0)\,\id x
\end{equation}
which coincides with $2I_0(\mu,j)$
 whenever $\nabla\cdot j=0$.
Using that extended functional to replace $I_0(\mu,j)$ in
\eqref{1rew}, we get
\begin{equation}\label{res2}
 I_F(\mu,j)= \frac 1{2}[G(\mu,j)- \caS_F(j)
  + H(\mu,F)]
\end{equation}
for all densities $\mu$ and stationary currents $j$.\\
We can still rewrite (\ref{res2}) in the form
\begin{equation}
 4I_F(\mu,j)=  \sup_F\bigl\{ 2\caS_F(j)-\si_F(\mu) \bigr\} - 2\caS_F(j)+\si_F(\mu)
\end{equation}
fully in terms of entropic quantities, due to
\eqref{ent-traffic}.\\  A similar canonical structure, cf.
\eqref{currentpotential} and \eqref{forcepotential}, has been
established already before in the framework of jump processes at
least on a sufficiently fine-grained scale of description, see
\cite{mn3}. One can therefore conclude that the structure of
(\ref{res2}) is also canonical in the sense of being generally
valid for a very large class of dynamics.

\subsection{Small fluctuations}
 We
look here at the Gaussian approximation in a dynamics far from
equilibrium. Later we will also make the driving $F$ small, to be
close to equilibrium.\\
 As is clear from (\ref{main2}), current and occupations are coupled. It is
because of this coupling that contractions of $I(\mu,j)$ to
$I(\mu)$ and to $I(j)$ become rather complicated. Even for small
fluctuations this coupling remains: take $\mu =
\rho(1+\epsilon\mu_1)$ and $j = j_{\rho}+\epsilon j_1$, with
$\epsilon$ a small parameter. Because $j-j_{\mu}$ is then
$O(\epsilon)$, the fluctuation functional is $O(\epsilon^2)$:
\begin{equation}\label{smagen}
\begin{split}
  I(\mu,j) &=  \frac{\epsilon^2}{4}\int \id x
  \bigl[ j_1 \cdot (\rho D)^{-1} j_1 + \mu_1^2 j_\rho \cdot (\rho D)^{-1} j_\rho
\\
  &\hspace{15mm}+ \nabla\mu_1 \cdot \rho D\nabla\mu_1
  - 2\mu_1 j_1 \cdot (\rho D)^{-1} j_\rho \bigr] + O(\epsilon^3)
\end{split}
\end{equation}
The last term in this approximation gives the coupling between
occupation and current fluctuations. It is proportional to the
stationary current, which is non-zero away from equilibrium. It is
only when we take a dynamics close to equilibrium, i.e. $F =
\epsilon F_1$, that the fluctuations decouple. In this
approximation we have that $j_{\rho} = O(\epsilon)$, and thus,
near equilibrium,
\begin{equation}\label{approx}
  I(\mu,j) =  \frac{\epsilon^2}{4} \int \id x \bigl[ j_1 \cdot (\rho D)^{-1} j_1
  + \nabla\mu_1 \cdot \rho D\nabla\mu_1  \bigr] +  O(\epsilon^3)
\end{equation}
with, to leading order, a complete decoupling between the
time-symmetric and the time-antisymmetric sectors.

\section{Contractions}

Now that we have a fluctuation functional for both symmetric and
antisymmetric variables, we can compute the statistics of
empirical averages of arbitrary physical quantities.  In
particular, we can try to find the fluctuation functionals for
density $I(\mu)$ and and for current $I(j)$  separately.  The
technique to do that is called contraction as we go so to speak to
a more contracted description. To start we look at the
fluctuations of the density (alone).

\subsection{Result 3: occupation statistics}\label{occ}

Look back at the definition \eqref{ocup-LD}.   As $I(\mu) = \inf_j
I(\mu,j)$, we have to compute the minimizing current $j$ for any
given density $\mu$. Since the minimization is constrained via the
stationary condition $\nabla\cdot j = 0$, we get the equation
\begin{equation}\label{eq-V}
 j = \chi\mu\bigl[ F-\nabla\cdot (U+\psi)\bigr] - D\nabla\mu
\end{equation}
where $\psi$ is a Lagrange multiplier (function of $x$). Not
surprisingly, we see that the minimizer is the stationary current
for a modified dynamics that makes $\mu$ stationary. This modified
dynamics is achieved here by replacing the imposed potential $U$
with a modified one, called $V = U+\psi$ (for some $\psi$). We
therefore call the minimizing current in \eqref{eq-V} $j_{\mu}^V$,
and the fluctuation functional becomes:
\begin{equation}\label{Imu-first}
  I(\mu) =  \frac{1}{4} \int (j_\mu^V - j_\mu)
  \cdot (\mu D)^{-1} (j_\mu^V - j_\mu)\,\id x
\end{equation}
For some explicit examples of solutions to~\eqref{eq-V},
see furtherdown, in equations~\eqref{V-circle}-\eqref{minep}.

The fluctuation functional $I(\mu)$ obtains other equivalent forms
by following a road backward from the one that led us before to
\eqref{main2}:
\begin{equation}\label{Imu-main}
  I(\mu) = \caT(\mu) - \caT_V(\mu) = \frac{\si(\mu) - \si_V(\mu)}{4}
\end{equation}
where the second equality follows again from~\eqref{ent-traffic}.
In this way we have recognized the excess traffic (or here also:
the excess entropy production) as governing the large time
statistics of the occupation times. That constitutes our third
main result: the fluctuation functional $I(\mu)$ exactly equals
one quarter of a difference in entropy production rates when
having density $\mu$, these rates being computed respectively for
the original dynamics and for a modified dynamics that makes $\mu$
stationary.\\

In formul\ae\ \eqref{Imu-first} and \eqref{Imu-main} the potential
$V$ has to be determined from $\mu$ by solving the inverse
stationary problem~\eqref{eq-V}. We now give two classes of
examples where this $V$ and hence $I(\mu)$ can be made explicit.

\emph{Diffusion on the circle.}
For the one-dimensional example of Section~\ref{sec: circle} the
inverse stationary problem~\eqref{eq-V} allows for an explicit
solution. The current $j_\mu^V$ is immediately read off the
formula~\eqref{circle-current},
\begin{equation}
  j_\mu^V = \frac{\be\overline{W}}{\int_0^1 (\mu D)^{-1}\id x},\qquad
  \overline{W} = \int_0^1 F\,\id x
\end{equation}
and the potential $V$ obtains the form
\begin{equation}\label{V-circle}
  V(x) = -\frac{1}{\be}\log\mu(x) + \int_0^x
  \bigl( F - \frac{j_\mu^V}{\be\mu D} \bigr)\,\id y
\end{equation}
which is a nonlocal functional of the given density $\mu$. The
fluctuation functional is explicitly given as
\begin{equation}
  4 I(\mu) = \si(\mu) -
  \frac{\overline{W}^2}{\int_0^1 (\mu D)^{-1}\id x}
\end{equation}
for $\mu\neq 0$.\\
 Observe that if $\mu = 0$ on some open set $A$
then the rate function equals $I(\mu) = \si(\mu)/4$. (That follows
also from the  equilibrium form \eqref{I-eq} below as the circle
gets \emph{effectively} cut and the dynamics mimics a detailed
balance one.) The infimum of $I(\mu)$ over all densities $\mu$
that vanish on $A$ then
gives the escape rate from the complement $A^c = [0,1) \setminus A$.\\
As a simple example, assume that $U = 0$ and let $F$ and $D$ be
some constants. In this case the entropy production (\ref{ent-prod}) reads
\begin{equation}\label{er}
  \si(\mu) = \be^2 D F^2 + D\int_0^1 \frac{\mu'^2}{\mu}\,\id x
\end{equation}
To compute the escape rate from $A^c$ (or, entrance rate to $A$)
we must take the infimum of \eqref{er} over all $\mu$ that vanish
on $A$.
 Setting $A = (0,\de)$ for some $0<\delta<1$, that infimum is reached
for the density $\mu^*(x) =
\frac{2}{1-\delta}\sin^2(\frac{\pi(x-\delta)}{1-\delta}), x\in
[\delta,1]$, and the escape rate is
\begin{equation}
  \inf_{\mu|_A = 0} I(\mu) = I(\mu^*)
  = \frac{\pi^2 D}{(1 - \de)^2} + \frac{\beta^2 D F^2}{4}
\end{equation}
Even in equilibrium ($F=0$) the result is meaningful as it relates
the diffusion constant to an escape rate.   In the context of
dynamical systems, the analysis of the escape rates and of their
link to linear transport coefficients was initiated by Dorfman
and Gaspard, see~\cite{Dorf,gasp} and references therein.

\emph{Close-to-equilibrium dynamics.} Let us start from what can
be said in general for equilibrium diffusions.  If $F = 0$ then
equation~\eqref{eq-V} has the solution $V = -\be^{-1}\log\mu$, and
the corresponding current $j_\mu^V$ and the entropy production
$\si_V(\mu)$ are both zero. As a result,
\begin{equation}\label{I-eq}
  I(\mu) = \frac{\si(\mu)}{4}
\end{equation}
This exact relation between the equilibrium dynamical fluctuations
and the entropy production is solely true for diffusion processes.
In contrast, for jump processes $\si(\mu)$ gives only the leading
term in an expansion of $I(\mu)$ around the equilibrium density
$\rho \propto e^{-\be U}$, and the relation \eqref{I-eq} obtains
corrections when beyond small fluctuations; see~\cite{dv1} for
details.\\

When close to equilibrium, the computation of $I(\mu)$ by
contraction is
    easy: we see from~\eqref{approx} that the second term on its right-hand side
     is
    just $I(\mu)$.\\
In the same approximation of small fluctuations and
close-to-equilibrium, the entropy production becomes:
\begin{equation}\label{approx_ent}
  \si(\mu) =  \int \id x \bigl[ j_{\rho} \cdot (\rho D)^{-1} j_{\rho}
  + \epsilon^2\nabla\mu_1 \cdot \rho D \nabla\mu_1 \bigr] + O(\epsilon^3)
\end{equation}
and thus we get
\begin{equation}\label{minep}
  I(\mu) = \frac{\si(\mu) - \si(\rho)}{4} + O(\ep^3)
\end{equation}
This reveals to be a special case of a general result, \cite{dv1},
according to which the entropy production governs the occupational
statistics in the linear irreversible regime. It provides a
fluctuation-based explanation for the minimum entropy production
principle introduced by Prigogine to characterize stationarity via
an (approximate) variational principle, \cite{prig}: the
stationary state has minimal (not necessarily zero) entropy
production.

\subsection{Current statistics}
 The contraction to the current $j$ is
also possible. However, up to special examples, there is no
explicit solution to the associated variational problem and for
general models one has to resort to a perturbative or numerical
analysis.  In fact, often the calculation starting from the
generating function of the current appears more practical than to
do the contraction starting from $I(\mu,j)$, see \cite{FNJ,der}.\\
 We restrict us here
to giving the result for a constant drift on the circle and to
small fluctuations around equilibrium.

\emph{Constantly driven diffusion on the circle.} Again we take
$U=0$ and $F,\chi$ constants. In this case, from~\eqref{main2} the
joined fluctuation functional reads:
\begin{equation}
 I(\mu,j) = \frac{1}{4D}\int\frac{1}{\mu}(j-\beta D F\mu -
 D\mu')^2\,\id x
\end{equation}
and for all $j$, the infimum over $\mu$ is reached at the uniform
distribution, so that
\begin{equation}
 I(j) = \frac{(j-\beta D F)^2}{4D}
\end{equation}
and hence we see that here the current fluctuations are Gaussian.

\emph{Close-to-equilibrium.} We have the analogue of the minimum
entropy production principle.  The starting point is
again~\eqref{approx} from which we extract the current
fluctuations:
\begin{equation}\label{maxi}
\begin{split}
  I(j) &= \frac{1}{4} \int (j - j_\rho) \cdot (\rho D)^{-1} (j - j_\rho)\,\id x + O(\ep^3)
\\
  &= \frac{1}{4} \bigl[ \scD(j_\rho) + \scD(j) - 2\caS(j) \bigr] + O(\ep^3)
\end{split}
\end{equation}
with $\scD(j) = \int j \cdot (\rho D)^{-1} j\,\,\id x$ sometimes
called the Onsager dissipation function, and $\caS(j)$ is the
entropy flux defined in \eqref{flu}. In particular, this leads to a
variational characterization of the steady current $j_\rho$ which
can be written as the following maximum entropy production
principle: the $j_\rho$ maximizes the entropy flux $\caS(j)$ under
the two stationary constraints
\begin{equation}
  \text{(1)  } \nabla \cdot j = 0,\qquad
  \text{(2)  }  \scD(j) = \caS(j)
\end{equation}
The second condition is indeed satisfied at $j = j_\rho$ (note
also that $\rho$ can with no harm in this order be replaced by the
equilibrium density $\rho_{F=0} = e^{-\be U}/Z$.) Such a
variational principle, known as a maximum entropy production
principle, is often used in applications and apparently even
beyond the linear irreversible regime.  As is however clear from
\eqref{maxi} from our dynamical fluctuation theory, the validity
of the maximum entropy principle is restricted to
close-to-equilibrium.  Beyond that regime, we must refer to
contractions from \eqref{main0}, \eqref{res2} or even from
\eqref{smagen} for generally valid expressions with a general
thermodynamic meaning.

\begin{acknowledgements}
K.N.\ acknowledges the support from the Grant Agency of the Czech
Republic (Grant no.~202/07/0404). C.M.\ benefits from the Belgian
Interuniversity Attraction Poles Programme P6/02. B.W.\ is an
aspirant of FWO, Flanders. C.M.\ and B.W.\ thank the Institute of
Physics of the Academy of Sciences in Prague for its kind
hospitality.
\end{acknowledgements}


\appendix

\section{Stochastic integrals}\label{stin}

We collect here some necessary albeit formal manipulations from
stochastic calculus. We refer to \cite{gard} and \cite{dif} for
further systematics.

Physical quantities, such as work and
heat dissipation, depend on the specific trajectory (or path)
that the system covers during its evolution.
But for diffusions these paths are not differentiable.
Therefore integrals like $\int_0^T f(x_t)\,\id x_t$ cannot
be defined in the usual way.
It turns out that these integrals,
called stochastic integrals, can be interpreted in different ways. Most common are the It\^o and Stratonovich
interpretations, see e.g.~\cite{gard}.

\emph{It\^o integral.} For the It\^o interpretation, the integral
domain $[0,T]$ is split up in a set of discrete points
$0=t_0<t_1<\ldots<t_n=T$, with $\Delta t_j = t_j-t_{j-1}$, such
that $\De t \equiv \max_j\Delta t_j\rightarrow 0$ for
$n\rightarrow\infty$. It is important to note that for
diffusions~\eqref{gsd1} we have that for $\Delta t_j\rightarrow
0$,
\begin{equation}\label{diffrel}
(\Delta x^{(\alpha)}_j)(\Delta x^{(\beta)}_j) = (x^{(\alpha)}_{t_j} - x^{(\alpha)}_{t_{j-1}})(x^{(\beta)}_{t_j} - x^{(\beta)}_{t_{j-1}})\rightarrow 2D_{\alpha\beta}(x_{t_j})\Delta t_j
\end{equation}
where $\alpha$ and $\beta$ denote the components of the vectors and the diffusion matrix.
The stochastic integral is then computed as
\begin{equation}\label{ito}
  \int_0^T f(x_t)\, \id x_t = \lim_{n\rightarrow\infty, \Delta t\rightarrow 0}
  \sum_{j=1}^n f(x_{t_{j-1}})\cdot(x_{t_j} - x_{t_{j-1}})
\end{equation}
Note that the function $f$ can be either scalar or vector, and it is evaluated at the left endpoint of the intervals
$[t_{j-1},t_j]$.

For the It\^o integral one cannot use the normal rules of
integration. Instead it is easily checked from~\eqref{diffrel} that for any function
$g$,
\begin{equation}
  \int_0^T \nabla g(x_t) \cdot \id x_t
  = g(x_T) - g(x_0) - \int_0^T (D\nabla \cdot \nabla g)(x_t)\,\id t
\end{equation}
(The symbol $\cdot$ stands for the scalar product.)

\emph{Stratonovich integral.}
The Stratonovich interpretation differs from the It\^o interpretation only in the points of evaluation of the function $f$. In this case $f$ is evaluated in the midpoints of the time intervals:
\begin{equation}\label{strato}
\begin{split}
  \int_0^T f(x_t) \circ \id x_t &= \lim_{n\rightarrow\infty, \Delta t\rightarrow 0}\sum_{j=1}^n f\bigl(\frac{x_{t_{j}}+x_{t_{j-1}}}{2}\bigr)\cdot(x_{t_j} - x_{t_{j-1}})
\\
  &=\frac{1}{2}\lim_{n\rightarrow\infty, \Delta t\rightarrow 0}\sum_{j=1}^n \bigl(f(x_{t_{j}})+f(x_{t_{j-1}})\bigr)\cdot(x_{t_j} - x_{t_{j-1}})
\end{split}
\end{equation}
where the symbol $\circ$ is commonly added as a notation to distinguish
between It\^o and Stratonovich interpretations. For our analysis
it is important to observe that the Stratonovich integral is
time-antisymmetric, and also that
\begin{equation}
  \int_0^T \nabla g(x_t) \circ \id x_t = g(x_T) - g(x_0)
\end{equation}

\emph{Relation between It\^o and Stratonovich.} For a scalar $f$, it is easily found from~\eqref{diffrel} that
\begin{equation}\label{str-relation}
  \int_0^T f(x_t) \circ \id x_t = \int_0^T f(x_t)\,\id x_t
  + \int_0^T (D\nabla f)(x_t)\,\id t
\end{equation}
(and analogously for vectors). Using that $x_t$ solves the
It\^o-stochastic equation~\eqref{gsd1}, that Stratonovich integral
further explicitly equals to
\begin{equation}\label{a7}
  \int_0^T f(x_t) \circ \id x_t =\int_0^T [f\chi(F - \nabla U) + \nabla \cdot (f D)](x_t)\,\id t
  + \int_0^T f(x_t) \sqrt{2D(x_t)}\,\id B_t
\end{equation}
Observe that the last It\^o-integral has mean zero since the
integrand evaluated at each mesh point $x_{t_{j-1}}$ and the
increment $B_{t_j} - B_{t_{j-1}}$ of the Brownian motion are
mutually independent, and the latter has zero mean. Hence, the
mean value of the Stratonovich integral~\eqref{str-relation} is
\begin{equation}\label{a8}
\begin{split}
  \Bigl\langle \int_0^T f(x_t) \circ \id x_t \Bigr\rangle_{\mu_0}
&= \Bigl\langle\int_0^T [f\chi(F - \nabla U) + \nabla \cdot (f D)](x_t)\,\id t\Bigr\rangle_{\mu_0}
\\
  &= \int_0^T \id t\, \int \mu_t \bigl[f \chi (F - \nabla U) + \nabla \cdot (f D)\bigr]\,\id x
\\
  &= \int_0^T \id t\, \int f\, \bigl[ \chi(F - \nabla U)\mu_t - D \nabla\mu_t \bigr]\,\id x
\end{split}
\end{equation}
The first equality is obtained by taking the mean of \eqref{a7}.
The second equality uses that $\mu_t$, defined by the
Fokker-Planck equation (\ref{fk1}), is the evolved density at time
$t$ when starting from $\mu_0$ at time zero.  Therefore $x_t$ is
there distributed according to $\mu_t$. The final equality is from
a partial integration.  That proves the
equality~\eqref{str-identity}.

\section{Path-distribution of diffusion process}\label{discretization}

The path of a particle subject to the stochastic
equation~\eqref{gsd1} is completely determined by the initial
condition and the realization of the Brownian motion $B_t$. Since
the latter is a standard Gaussian process, we can integrate out
the noise to obtain a path-integral representation. We restrict
here to a simple heuristic argument which goes as follows.

To find the probability for the particle to pass through (the
neighborhood of) a discrete set of points $\om = (x_0,t_0 = 0;
x_1,t_1; \ldots; x_n,t_n = T)$, and having
 already in mind the limit $n \rightarrow \infty$, we use that each increment $\De x_j = x_j - x_{j-1}$ is by~\eqref{gsd1} in a one-to-one correspondence with an increment of the Brownian motion, which is
$(2D(x_{j-1}))^{-1/2}[\De x_j - (\chi F - \chi\nabla U + \nabla
\cdot D)\,\De t_j]$. Since the latter increments are independent
standard Gaussian variables, we immediately obtain the discrete
path-distribution (or its density with respect to the
$(n+1)-$product of flat distributions):
\begin{multline}\label{path-integral}
  \id\bsP_{\mu_0}(\om) \simeq \mu(x_0)\, \de x_0 \prod_{j=1}^n [4\pi D(x_{j-1})\, \De t_j ]^{-\frac{1}{2}}
\\
  \times \exp \Bigl\{ -\frac{1}{4 \De t_j}
  \bigl[\De x_j - (\chi F - \chi \nabla U + \nabla \cdot D)(x_{j-1})\, \De t_j \bigr]
\\
  \cdot
  D^{-1}(x_{j-1}) \bigl[\De x_j - (\chi F - \nabla U + \nabla \cdot D)(x_{j-1})\, \De t_j \bigr]
  \Bigr\}\,\de x_j
\end{multline}
where the initial point $x_0$ is sampled from an initial distribution $\mu_0$.
A useful representation employs the reference process
\begin{equation}\label{ref}
  \id x_t = \nabla \cdot D(x_t)\,\id t + \sqrt{2 D(x_t)}\,\id B_t
\end{equation}
with the initial point sampled from the flat distribution; note
the latter (unnormalizable in general) distribution is stationary.
Its path-distribution $\bsP^0$ is simply obtained
from~\eqref{path-integral} by putting $F = \nabla U = 0$ and $\mu
= 1$. Comparing both path-distributions and passing to the limit
$n \rightarrow \infty$, $\De t \rightarrow 0$, we obtain a
Girsanov-formula for diffusions:
\begin{multline}\label{b3}
  \frac{\id\bsP_\mu}{\id\bsP^0}(\om) = \mu(x_0)\,\exp \Bigl\{
  -\frac{\be}{4} \int_0^T [(F - \nabla U) \cdot \chi(F - \nabla U)](x_t)\,\id t
\\
  -\frac{1}{2} \int_0^T [(\nabla \cdot \chi) \cdot (F - \nabla U)](x_t)\,\id t
  + \frac{\be}{2} \int_0^T (F - \nabla U)(x_t) \cdot \id x_t \Bigr\}
\end{multline}
where an It\^o integral comes out by construction. Using the
relation~\eqref{str-relation} to replace that integral with a
Stratonovich one, we finally arrive at
formulas~\eqref{path-distribution}--\eqref{lagr-minus}. A rigorous
derivation can e.g. be found in~\cite{LS}.



\begin{thebibliography}{10}
\bibitem{rem}
More precisely, that asymptotic law needs to be understood in the sense
$\lim_{\ep \to 0} \lim_{T \to \infty} (1/T) \log\bsP(\bar\mu_T \stackrel{\ep}{=} \mu) = I(\mu)$, where
$\bar\mu_T \stackrel{\ep}{=} \mu$ stands for a neighborhood of $\mu$ in a weak metrizable topology (see e.g.~\cite{DZ}, appendix D.2 for more details).
\end{thebibliography}
\end{document}